# Reply to a review of photon-exchange interactions by Opatrny and Kurizki

J. D. Franson
Johns Hopkins University
Applied Physics Laboratory
Laurel, MD 20723 U.S.A.

Opatrny and Kurizki [1] have reviewed a number of issues related to the feasibility of quantum computation based on photon exchange interactions. Most of the issues that they considered have already been discussed in more detail in our earlier publications. For example, my brief discussion of collisions in Ref. [2] was followed by a more detailed density matrix analysis [3] that the authors did not mention. I would like to review each of these issues and then summarize my view of the overall prospects for this approach to quantum computing.

## I. SCATTERING AND DISPERSION

The authors have argued that the emission of photons into other modes of the electromagnetic field must invalidate the results of our analysis. We have considered two different limiting cases, for which the effects of scattering and dispersion are different: (a) Weak coupling and large detunings and (b) Strong coupling on or near resonance.

(a) Weak coupling and large detunings. In order to simplify the analysis, most of our early work considered the limiting case of a weak interaction between two photons and the atoms in a thin medium, with the photons detuned far from the resonant frequency of the atoms. Although the conditional phase shifts produced in this way are too small to be of any practical use, they were intended to illustrate the origin of the nonlinear phase shift. All of Ref. [2] and most of Ref. [3] correspond to this situation.

It is an experimental fact that scattering and dispersion are small in the limit of large detunings and it seems reasonable to simply neglect them in an analysis to a first approximation. This is equally true for the incident wave packets that we considered in Ref. [3] as it is for plane waves. Scattering into other field modes will primarily have the effect of depleting the probability amplitude of the original state with no significant effect on the coherent process of interest, in which both photons emerge in the same state that they entered the medium aside from the nonlinear phase shift. In particular, scattering into other field modes has no effect at all on the nonlinear phase shift of interest to lowest order in perturbation theory. That is not to say that scattering will not occur or that it will not be larger in magnitude than the phase shifts of interest, but it does mean that scattering into other modes can be neglected in calculating what the nonlinear phase shift is. Once again, these small phase shifts in the weak-coupling limit were not meant for practical applications but only as a way of understanding the origin of the nonlinear effect as simply as possible.

(b) Strong coupling on or near resonance. In the latter part of Ref. [3], we addressed the fact that strong interactions will be required in any practical devices, in which case the emission of photons into other modes of the field cannot be ignored. We briefly stated that, in the case of strong coupling, it should be possible to use momentum conservation in a thick medium to inhibit emission into extraneous modes of the field compared to the rate of re-emission back into the original modes. We subsequently performed [4] a detailed analysis of such a situation using the Dicke-state formalism, which showed that emission back into the original field modes will be enhanced by a factor of $\sqrt{N}$ compared to spontaneous emission into other modes, where $N$ is the number of atoms in the medium. Aside from technical issues such as the required optical properties of the medium, which I will address below, this demonstrates that emission into other field modes can be neglected even in the strong-coupling limit, as we suggested in Ref. [3].

To summarize the scattering issue, the emission of photons into extraneous modes of the field can be neglected in the limit of weak coupling and it can be minimized in the strong-coupling case using the techniques described in Ref. [4].

## II. COLLISIONS

As I showed in Ref. [2], all of the Feynman diagrams for the process of interest will cancel each other unless the medium is perturbed in some way, and I gave an intuitive argument to suggest that collisions of the atoms with a buffer gas would affect some of the diagrams more than others and eliminate the cancellation. The situation is similar in some respects to the collision-induced resonances observed by Bloembergen and his colleagues [5], where the cancellation of diagrams is also removed by collisions.

Following the publication of Ref. [2], we performed a more detailed density matrix calculation to investigate the effects of collisions in this system [3]. It was found that collisions would eliminate the cancellation of the diagrams and give a nonlinear phase shift under the same assumption made in Ref. [2], namely that all of the virtually-excited states dephase at the same rate. We also pointed out that this assumption is non-physical, but that nonlinear phase shifts would be obtained if $\tau_1 \neq \tau_2$, where $\tau_1$ and $\tau_2$ are the dephasing rates of virtually-excited atomic states produced by the absorption of two photons with different frequencies $\omega_1$ and $\omega_2$. Finally, we stated in Ref. [3] that the decoherence produced by the use of collisions in this way is sufficiently large that the approach did not appear to be of any practical use.

Although the authors have not raised the issue, it has previously been suggested that it may not be possible for $\tau_1$ and $\tau_2$ to have different values even when the photon detunings are different. To see that this can occur near a resonance with an inelastic collision process as we suggested, consider the three-level atom shown in Figure 1. The virtual absorption of photon 1 or 2 can produce a transition between levels 1 and 2, while collisions can produce a transition between levels 2 and 3; level 3 is assumed to have a negligible radiative decay time. In order to be specific, I will assume that levels 2 and 3 are coupled by a dipole-dipole interaction of the form $\alpha \mathbf{r}_a \cdot \mathbf{r}_b / |\mathbf{R}_a - \mathbf{R}_b|^n$. Here $\alpha$ is a constant, $\mathbf{r}_a$ and $\mathbf{r}_b$ are the relative coordinates of one of the atoms of interest and a buffer-gas atom, respectively, and the value of $n$ will depend on whether the dipole moments are permanent or induced. If the center-of-mass locations, $\mathbf{R}_a(t)$ and $\mathbf{R}_b(t)$, of a pair of atoms are treated classically, then the Hamiltonian will contain a time-dependent term that depends on the separation of the atoms. Assuming uniform velocities for the two atoms and a distance $r_0$ of closest approach, Schrodinger's equation can be integrated numerically to calculate the time evolution of the state vector. Figure 2 shows the phase shift produced in the probability amplitude of a virtually-excited atomic state (level 2) as a function of the detuning of the photons for a nominal choice of parameters. It can be seen that the magnitude of the phase shift in a virtual atomic state due to a collision does depend on the photon detuning, which is consistent with the assumption that $\tau_1 \neq \tau_2$.

To summarize, collisions can produce a nonlinear phase shift but the decoherence rate is too large for practical applications, as we previously stated in Ref. [3].

### III. PULSE SEQUENCES

The ability to produce conditional (nonlinear) phase shifts in our approach is based on the fact that exchange interactions will cause two photons propagating in a medium to be in an entangled state that we referred to as a two-photon dressed state [3]. If $|\psi_b\rangle$ represents the state of the system when both photons are present in the medium at the same time and $|\psi_1\rangle$ and $|\psi_2\rangle$ represent the states of the system when only photon 1 or 2 is present, then we showed that

$$|\psi_b\rangle \neq |\psi_1\rangle |\psi_2\rangle \tag{1}$$

For example, the pulse sequences described in the preprint of Ref. [4] can produce the entangled state

$$|\psi_b\rangle = (|0\rangle + |\gamma_2, \gamma_2\rangle)/\sqrt{2} \tag{2}$$

Here $|0\rangle$ represents a state with no photons and two virtually-excited atoms, while $|\gamma_2, \gamma_2\rangle$ represents a state with two photons of frequency $\omega_2$ and no excited atoms. It is generally recognized that linear optical elements, such as beam splitters, cannot produce an entangled state. As a result, Equation (2) suggests that nonlinear effects can be produced at the two-photon level using photon-exchange interactions.

We analyzed the case of a single laser pulse in detail in Ref. [3] and showed that it can produce a nonlinear phase shift whose origin can be intuitively understood. On the other hand, a single laser pulse will also leave the system in a state that is somewhat different from the input state, which would correspond to a nonzero probability for an error in quantum computing applications. What is required is a sequence of pulses that will produce a nonlinear phase shift of $\pi$ while leaving the system in its original state at the end of the sequence. Designing such a pulse sequence is a challenging task, since the fact that the system evolves differently when both photons are present makes it difficult to leave the system in its original state regardless of whether one or two photons are present. The design of pulse sequences of this kind is analogous in some respects to the design of NMR pulse sequences, which is often done using computer programs with limited physical insight into the details of an actual sequence.

At this point, we have considered a number of potential laser pulse sequences with varying degrees of success. The "five-pulse" sequence described in Ref. [3] assumed that the laser pulses had negligible bandwidth, in which case they could be tuned to excite any desired two-photon dressed state. Contrary to the comments of the authors, Ref. [3] considered all of the relevant states, including the state with two photons in mode 2. In principle at least, a sufficiently narrow-band pulse can be tuned to avoid the excitation of the latter state, since the dressed-state energy differences between the relevant states are not equally-spaced as they are in free space. These energy differences are due to conventional mechanisms and are sufficiently small that this approach does not appear to be of practical value, as we have already stated in Ref. [4].

M. Lukin [6] previously noted that the "three-pulse" sequence in the preprint version of Ref. [4] was incorrect for the same reasons that the authors describe here. In essence, the phase shift experienced by photon 1 when it is present alone was neglected on the basis that it is the same regardless of which path it follows in the usual interferometer implementation of an optical quantum logic gate; that argument is not correct for the reasons pointed out by Lukin [6] and by the authors.

We have also described [3] a "ten-pulse" sequence that the authors did not mention. Aside from the fact that it was computer-designed and is complex, it also requires the assumption that the Stark shift from a laser pulse can produce phase shifts $\Delta\theta_1$ and $\Delta\theta_2$ in level 2 of an atom that has been virtually excited by photons 1 and 2, respectively, with $\Delta\theta_1 \neq \Delta\theta_2$ due to the difference in the photon detunings. The situation here is essentially the same as for the three-level atom shown in Figure 1 if levels 2 and 3 are assumed to be coupled by a laser pulse instead of a collision process. A calculation similar to that described above for the case of collisions also gave the result that $\Delta\theta_1 \neq \Delta\theta_2$ when the photon detunings are different, in agreement with the assumption made in Ref. [3]. This pulse sequence gives a small nonlinear phase shift and may be too complex for practical applications, as we stated in Ref. [3].

We are currently investigating other sequences of laser pulses that are also based on the use of Stark shifts in an effort to find a more optimal sequence of pulses. The additional atomic state (level 3) required for the production of Stark shifts is now included in the state vector, which eliminates the need for any simplifying assumptions with regard to the effects of the laser pulses. The cooperative nature of the Dicke states can have an effect on the Stark shifts in addition to eliminating the emission of photons into other field modes, as will be described in the revised version of Ref. [4]. We expect that the complexity of the required pulse sequences will not be substantially greater than that required for the pulses (laser or electrical) used in other approaches, such as NMR or ion traps.

## IV. OVERALL PROSPECTS FOR PHOTON-EXCHANGE INTERACTIONS

Before discussing the potential advantages of our approach, let me review some of the challenges that will have to be met before an operational quantum logic gate can be constructed. First of all, the medium will have to have high density, a large oscillator strength, and phonon-free lines to minimize decoherence. Other groups [7] are currently investigating newly-developed materials with properties of that kind for similar purposes. I agree with the authors that the devices will probably have to be cryogenically cooled, but I do not view that as a significant difficulty. I would guess that inhomogeneous broadening will eventually be the limiting factor with regard to decoherence in these devices, and that this will determine the required speed of the logic operations. The authors' numerical estimates do not appear to reflect the performance that may be achievable using new materials and the approach outlined in Ref. [4].

The approach that we have outlined [4] is well suited for the operation of quantum logic devices in which the photons propagate in waveguides on the surface of an optical medium. As a result, it may eventually be possible to fabricate large numbers of logic gates and memory devices on the surface of a suitable crystal, in analogy with current semi-conductor techniques. In addition to the material issues mentioned above, it will also be necessary to develop suitable techniques for the fabrication of low-loss optical waveguides for this purpose.

Any optical approach to quantum computing will require an efficient source of true single photons to initialize the computation. We have previously suggested [4, 8] that the quantum logic gates themselves could be used to perform quantum non-demolition measurements on a large number of optical fibers or waveguides containing an average of one photon each, as generated by a classical laser pulse; those channels found to contain one and only photon could then be switched into the input of the computer. We have also suggested [4, 8] that quantum non-demolition measurements could be used as a way of enhancing the single-photon detection efficiency required in order to read the output of an optical computer. The same waveguide structures could be used for memory or logic operations, since a single photon could be stored in the medium using a $\pi$ pulse and then restored on demand using a second pulse. It appears that all of the necessary functions for quantum computing can be performed provided that the basic logic gates are sufficiently efficient.

Assuming that these challenges can be met, our approach would have a number of potential advantages for the construction of full-scale quantum computers, or for the less ambitious goal of implementing quantum repeaters for use in quantum cryptography. An optical approach has the advantage that optical fibers or waveguides could be used to make connections between any two logic gates or memory devices. Most other approaches are dependent on nearest-neighbor interactions and would not allow the use of the "bus" architecture of conventional computers. Optical approaches would also have the advantage of high speed operation, which may be a consideration if many logic operations are required in order to perform a useful computation. Our optical approach has the additional feature that it does not require the use of optical cavities or atomic traps, which would be an advantage in the construction of a full-scale quantum computer containing a large number of qubits.

In summary, there are a number of challenges that would have to be met in order to develop efficient quantum logic gates using photon-exchange interactions, including issues related to materials, fabrication, and efficient pulse sequences. If those challenges can be met, our approach would have a number of potential advantages for the eventual implementation of a quantum computer.

## ACKNOWLEDGMENTS

I would like to thank M. Lukin and T. Pittman for their comments. This work was supported by ONR, ARO, NSA, and NASA.

___

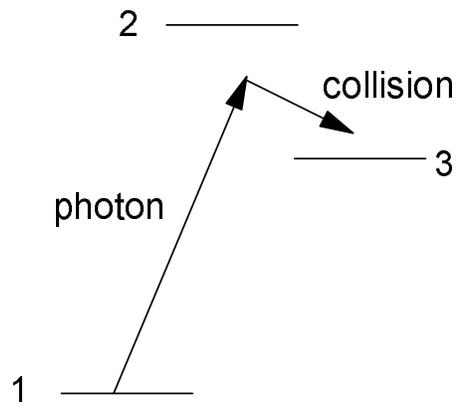

Fig. 1  A three-level atom in which the absorption of a photon can produce a virtual transition from level 1 to level 2, after which a collision can produce a virtual transition from level 2 to level 3 via a dipole-dipole interaction.

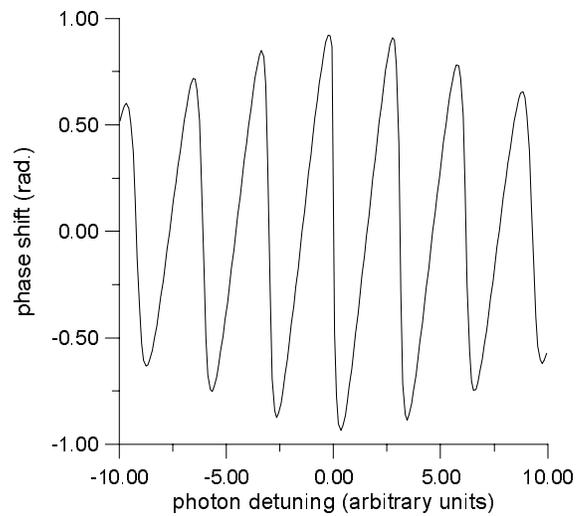

Fig. 2.  Phase shift produced in the probability amplitude of a virtually-excited atomic state (level 2 of Fig. 1) as a function of the detuning of the photon.